# Phase Transitions of the First Kind as Radiation Processes

Mark E. Perel'man [a], Vitali A. Tatartchenko [b,*].

[a] *Racah Institute of Physics, Hebrew University, Jerusalem, 91904, Israel*
[b] *Saint-Gobain Crystals, France*

[*]*E-mails*: m.e.perelman@gmail.com; vitali.tatartchenko@orange.fr

**ABSTRACT.**

Crystallization and vapor condensation are considered as processes of sequential entering of single atoms/molecules into condensate. The latent heat can be carry away by radiation of characteristic frequencies generated in the course of transition. The estimated dependences of latent (radiated) energy of boiling on temperature confirm and prove the well-known empirical Trouton's rule applicable to many simple substances. It leads to the estimation of interrelation of critical parameters of corresponding substances. Experimental results of the authors and other researchers concerning crystallization from the melt of different substances (alkali halides, sapphire, tellurium, ice, copper) are presented, as well as condensation of water vapor, the correspondence to the offered model is established. It allows developing of "the spectroscopy of phase transitions", and can lead to control of crystallization process, to crystallization stimulated by the characteristic radiation, etc. Formation of clouds in our atmosphere should be accompanied by characteristic radiation detectable for meteorological warnings.

## 1. INTRODUCTION

In the paper [1] we had briefly considered phase transitions of the first order into more ordered states as a set of single atoms/molecules entering into condensate with removing of latent heat by the transitive (Ginzburg-Frank) radiation. Here we try to give unwrapped theoretical considerations, consider their consequences and analyze experimental data.

The initial base for our consideration is such. All peculiarities of phase transitions should be definable, in principle, by electromagnetic properties of constituents. So, the basic parameter of the first-order phase transition, its molar latent heat $\Lambda(T)$ should be determinable, undoubtedly, via electromagnetic parameters of particles and substances as a whole and via extensive parameters, the temperature T, pressure p, etc. In the simplest case of particles with zero electromagnetic parameters (atoms of noble gases), the unique electromagnetic interactions should arise only via vacuum fluctuations resulting in the Casimir forces. In the theory [2] as well as further in [3] it was shown that such approach leads to the satisfactory estimation of the latent heat of Helium liquefaction.

For particles possessing certain multipole moments, the consideration of phase transitions should be carried out with taking into account, at least, their most intensive forces. It may be supposed that the removal of liberated latent heat or its greatest part, at least, i.e. the energy generated during transition, can be examined as a key to such consideration, since the possibilities and channels of the removal processes can indicate on changes of constituents' interactions during the transition processes.

However, it is apparently tantamount to an implicit assumption that the removal of latent heat (bond energy) at transition into more condensed phases should occur via heat conductivity, even at constant temperature in the condition of supersaturation, etc. So, possible ways of the removal of latent heat are leave out even under the kinetic approach to condensation and solidification processes, which describe them as a sequence of transitions of separate atoms/molecules through phase borders (e.g. [4]).

But the problem is not so simple even logically: heat conductivity depends on a difference of temperatures and is executing as a continuous process, but phase transition can be a set of discrete processes at the constant temperature. Indeed, let's consider for its specification a process of vapor condensation (sublimation) in a rarefied substance where heat conductivity is complicated or is in general impossible: e.g. at a fast and homogeneous fog formation in an atmosphere or processes of gas condensation in cooling planetary clouds. At such non-equilibrium transitions the removal of latent heat by a non-Planckian radiation is only possible; its spectrum should be determinable by quantum numbers of initial and final states and therefore must contain characteristic frequencies.

Notice, that the imperative interrelation of the latent heat generated in the phase transition of the first kind with optical characteristics of forming phases can be predictable at consideration of so called residual rays [5] or dielectric anomalies [6]: the factor of reflection aspires to the unity for frequencies corresponding to the latent heat of many crystals, more often to the half of its magnitude (cf. [7]). It is necessary to add here that the change of symmetry, e.g. at the crystal growth in vapor phase, requires an emission of scalar particle, so called Goldstone boson [8] that would be noticeable in the spectra of removing latent energy. It can be added also that opposite processes, e.g. melting, can be executed without phonons at all, by direct electron-photon interactions [9].

The necessity of investigations of removal problems have been put forward and elaborated, in the frame of quantum electrodynamics, in the papers [10, 11] [1).

Generalization of such approach requires the microscopic consideration, i.e. the kinetic approach to processes of phase transitions based on the examination of electromagnetic characteristics of atoms/molecules in both phases. The consideration of elementary processes of particles joining with condensate should lead to possibilities of estimations of latent heat and to revealing its dependence on temperatures of transition. This problem can be complicated, in principle, by clusters formation in a gas phase and their subsequent entrance as peculiar particles into a condensed phase; these opportunities are not completely examined here.

The interrelation of molar latent heats $\Lambda_b$ and temperatures of boiling $T_b$ was empirically established in 1884 by F. T. Trouton for many substances (non-polar or weakly-polar) under normal atmospheric pressure [12] with an accuracy of the order of 7%:

$\Lambda_b/T_b \sim 21.3$ cal/mole K        (1.1)

This rule and several its refined variants represent the base for many physicochemical estimations, the history and further refinements of the Trouton's rule are described e.g. in [13]. However, till now this rule is substantiated only thermodynamically, but not in the frame of microscopic theory, therefore its physical sense and significance remain unknown. (The consideration in [10, 7] reveals its physical sense, but in a general form only and does not lead to possibilities of its immediate refinements.)

---

[1]). At the discussion of articles [10, 11] A. D. Sakharov noted (1971) that an existence of such emission corresponds to common dimension principles: the liberation of latent energy can go sometimes, at least, as a volumetric process and therefore can not be limited always by the heat conductivity via the two-dimensional surface: the volumetric ways of its releasing, the processes of a non-Planck radiation, must exist as well.

We shall attempt to consider the process of transition into more ordered state microscopically as the conversion, completely or partially, of generated latent heat into the electromagnetic transitive radiation [14] (the general theory of transitive radiation in [15]). The intensity and spectrum of such phase radiation should depend on multipole moments of particles in both substances and on features of dispersions of $\varepsilon$ and $\mu$.

Note, that in the article [16] the possibility of radiation is estimated for phase transitions in solids accompanied by motion of charges from one equilibrium position to another. A ferroelectric transition of the first type is an example of these phase transitions. In this case, the sample polarization changes from some fixed value to zero during the transition and the radiation has to be located in the range of $\lambda \sim 1$ cm. Such approach can has, as it seems, only restricted value and can be applied to certain definite processes.

Calculations of intensity of the radiations induced by electromagnetic moments (we consider for simplicity the dipole radiation) are carried out in the Section 2 at the assumption that phase transition occurs as a one-stage process, without additional adsorption phenomena (cf. e.g. [17]), to the arguments given in [1] several thermodynamic arguments are here added. However, for further development the representation of phase transition in terms of interaction of a condensable particle with its "mirror" reflection in medium, i.e. the "recombination" or neutralization of anti-parallel dipoles, is useful (Section 3). Such representation reduces the problem to examination of dipole - anti-dipole system, i.e. to the QED problem, and allows the estimating of the main peculiarities of spectra of this radiation (Section 4).

In the Sections 5-7 are consequently considered experimental data related to infra-red transparent substances, water and infrared nontransparent substances and their comparisons with theoretical predictions are executed. Since we are interested in the principal physical consideration, we will restrict the comparison of calculated forms with the simplest empirical relation (1.1) only (more complicated and specialized expressions can be considered elsewhere). As the spectrum of such non-equilibrium phase radiation is non-Planckian, it must be determinable by establishment of new bonds at transition of single particles into a new phase. The partial numerical consideration of individual substances is not needed since a lot of them satisfy the Trouton's rule.

The results and certain perspectives are summed in the Conclusions.

## 2. RADIATIVE TRANSITIONS

If medium contains particles possessing the electric dipole moment, the phase transition of the first kind can be considered as an instantaneous change at the time t = 0 (or in a point z =0) of the dielectric susceptibility of substance and also, probably, the value of dipole moments and positions of particles:

$$\varepsilon(t) = \varepsilon_1\theta(-t) + \varepsilon_2\theta(t); \quad \mathbf{d}(t) = \mathbf{d}_1\theta(-t) + \mathbf{d}_2\theta(t), \qquad (2.1)$$

where $\theta(t)$ is the Heaviside step function. (It is accepted, for simplicity, that $\mu = 1$; in more general case the impedance $Z = (\varepsilon/\mu)^{1/2}$ must be considered.)

The intensity of transient radiation can be calculated directly. But these direct calculations are not necessary, since they can be simply estimated by the known analogy between transient radiations of (nonrelativistic) charge q and dipole $\mathbf{d}$ [15]. So, if the charge q is entered, under an angle $\theta$ and with a velocity $\mathbf{v}$ from vacuum into the medium, the intensity of transient emission is

$$W(\omega,\theta) \approx \frac{2q^2v^2}{\pi c^3}\left|\frac{\varepsilon-1}{\varepsilon}\right|^2 \sin^2\theta. \tag{2.2}$$

After averaging over angles it can be assumed that for the transition from radiation of charges to radiation of dipoles such replacement in (2.2) is enough:

$$q^2 \to \tfrac{1}{4}(\mathbf{d}_2^2 + \mathbf{d}_1^2)(\omega/v)^2 \approx \tfrac{1}{2}\mathbf{d}^2(\omega/v)^2. \tag{2.3}$$

Let's specify the expression (2.2) by substitution the usual magnitude $\varepsilon(\omega)$ with only one, for simplicity, resonant frequency:

$$\varepsilon(\omega) = 1 + \omega_P^2/[\omega_0^2 - (\omega + i\Gamma)^2] \approx 1 + \omega_P^2/2\omega_0[(\omega_0 - \omega) - i\Gamma/2], \tag{2.4}$$

where $\omega_P = (4\pi N r_0 c^2)^{1/2}$ coincides with the plasma frequency, N is the density of scatterers, $r_0$ is the classical radius of an electron with the mass of free particle.

In view of the subsequent integration

$$|\varepsilon(\omega) - 1|^2 = \omega_P^4/4\omega_0^2[(\omega_0 - \omega)^2 + \Gamma^2/4] \to (\pi\omega_P^4/2\omega_0^2\Gamma)\,\delta(\omega_0 - \omega). \tag{2.5}$$

Integration of (2.2) over angles and frequencies with the help of (2.5) leads to the magnitude of energy radiating at the entering of single dipole into condensate:

$$W^R \approx \omega_P^4 d^2/6c^3\Gamma = \pi(c/\Gamma)\,\sigma_T\, d^2 N^2, \tag{2.6}$$

where d is the mean value of dipole moments in both substances, $\sigma_T$ is the Thompson cross-section. The characteristic dependence $N^2$ shows that (2.6) represents the second term of decomposition of the complete energy in substance W(N) in terms of particles density describing the pair interactions.

For indicative numerical estimations can be taken on $|\mathbf{d}| \sim 1\,D = 10^{-18}$ CGSE, density of particles of substance $N \sim 3\cdot 10^{22}$ cm$^{-3}$ and $\Gamma \sim 10^8$. From here an estimation $W^R \sim 5.6\cdot 10^{-13}$ erg/particle or a latent molar heat of transition $\Lambda = N_A W^R \sim 33$ kJ/mole = 8 kcal/mole follow under normal temperature and pressure. Such estimations generally correspond to the order of magnitudes of observable values and support our assumption about possibilities of removing of latent heat by radiation.

Let's consider the physical sense and possible interpretations of (2.6). The factor $d^2/\Gamma$ can be considered as the imaginary part of polarization of medium: $d^2/\Gamma \to \hbar\,\mathrm{Im}\,\alpha$. Hence (2.6) can be expressed via the complex susceptibility $\varepsilon(\omega) = 1 + 4\pi\alpha(\omega)N$ as

$$W \leq \tfrac{1}{4}\hbar c\sigma_T N\,\mathrm{Im}\,\varepsilon(\omega_0). \tag{2.7}$$

Thus, the developed approach predicts presence of peculiarities in spectra of the condensed phase connected with emitted frequencies.

Let's rewrite (2.6) via durations of elementary processes. So, inasmuch as $\tau_\gamma = 1/\Gamma$ can be considered as the "duration" of emission process, and the parameters $\ell = 1/\sigma_T N$ and $\tau_0 = \ell/c$ can be considered as a length and duration of photons' free path up to interaction with other constituents of medium [18], the main expression (2.6) can be rewritten as

$$W^R \approx \pi(\tau_\gamma/\tau_0)W_N, \tag{2.8}$$

where $W_N = d^2 N$ is the energy of dipoles interaction. Thus, if $\tau_\gamma < \tau_0$ the energy of bonds can be radiated by atom/molecule at its entry into condensate, till reabsorption by another center (compare the consideration in [7], the joining of particles may lead to change of these parameters including the Dicke "superradiation", such example is considered in [19]). So, this condition can be expressed as the restriction on the magnitude of effective level width in the condensate, for which radiative transitions can be possible:

$$\Gamma > \Gamma_{thr} = c\sigma_T N. \tag{2.9}$$

At substituting of this threshold value in the general expression (2.6), i.e. at assuming that $\tau_\gamma = \tau_0$, the value of threshold condensation energy of particles having dipole moment follows:

$$W^R_{thr} \approx \pi d^2 N. \tag{2.10}$$

Let's refer to the thermodynamic description. By accepting for polarizability of dense substances $\alpha \sim 1/N$ and taking into account that at the same time for the case of orientation polarizability $\alpha \approx d^2/3\kappa_B T$, the expression (2.10) can be converted to

$$W^R_{thr} \approx 3\pi\kappa_B T, \tag{2.11}$$

which is equivalent to the universal expression for molar latent energy of the condensed phase formation $\Delta H_{form} = N_A W^R_{thr}$ consisting from particles with dipole moments:

$$\Delta H_{form}/T \approx 3\pi R_{gas} = 78.2 \text{ J/mol K}, \tag{2.12}$$

where $R_{gas}$ is the universal gas constant.

This result proves the universal correlation of energy of the condensed state formation for dipole particles and temperatures of transition at $T \ll T_c$. Hence, it ascertains that the potential energy of bonds of simple particles at (normal) temperature of phase transition into the condensed phase is $2\pi$ times bigger their average kinetic energy.

The numerical evaluations of (2.12) practically coincide with the Trouton's rule (1.1) for latent energy of vapor condensation: $\Delta H_{cond}/T \approx 80$ J/mole K. Some distinction between energies of formation and boiling can be connected to certain simplifying guesses and approximations of our estimations and does not belittle the general importance of this correlation.

## 3. RECOMBINATION OF THE COUPLED ANTI-PARALLEL DIPOLES

If the particle of condensable gas possesses certain multipole moment, constant or virtual, at the approach to a surface it induces the redistribution of charges in medium, which is describing as the occurrence of a "mirror" moment (charge dislocation) in its near-surface adsorption layer. The arising attraction accelerates the sedimentation of condensable particles on the surface of condensate, i.e. their adsorption and/or phase transition.

Let's consider, as well as above, the condensation of vapor atoms or sufficiently simple molecules with the dipole moment **d** on a flat surface of condensate at constant temperature much lower critical and at such pressure that the vapor can be considered as an ideal gas. Kinetic energies of particle and of its image are equal to $U_K = \frac{3}{2} \kappa_B T$; the energy of interaction of two identical dipoles on distance **R**, averaged over angles, is equal to $U_R = \mathbf{d}^2/R^3$.

According to the virial theorem (e.g. [20]) the "temperature" part of potential energy of system, potential of which depends on distance as $\mathbf{R}^n$, is $U_T = -(2/n)U_K$, i.e. in our case $|U_T| = \frac{2}{3} U_K \equiv \kappa_B T$ for each dipole. Inasmuch as at entering of such particle into condensate its temperature should not vary, i.e. the kinetic energy and corresponding part of potential energy should be kept, the liberated energy (the latent heat generated in the first-order phase transition of single particle) should be equal

$$\Delta U_b \equiv U_R - |U_T| = \mathbf{d}^2/R^3 - 2\kappa_B T_b. \tag{3.1}$$

In this expression, obviously, dynamic polarizability of atom, retardation and Casimir forces are not taken into account. Their role can be more significant at consideration of completely neutral particles (atoms of noble gases, molecules $CO_2$, $CH_4$, etc.). It is possible to notice that, apparently, just this circumstance essentially lowers their temperatures of phase transitions in comparison with atoms/molecules possessing such moments (atoms of metals, at which the S-state is basic, e.g. Lithium, can form dimers and more complex aggregates with diversified moments in a gas state).

Feasibility of the condition (3.1) is determinable, certainly, by an opportunity of approximation of substance polarization via only one "mirror" dipole. Hence the condensable dipole must be completely outside of condensate, i.e. the distance R in (3.1) should exceed the sizes of particle. Therefore this expression can be unfit or too complicated for consideration of big molecules.

The expression (3.1) can be considered, within the framework of described restrictions, as the strict one. In substances with orientation polarizability (e.g. [21]) $\alpha \approx \alpha_0 + \mathbf{d}^2/3\kappa T$ and therefore it somewhat specifies (2.11):

$$\Delta U_b/\kappa_B T_b = 3(\alpha - \alpha_0)/R^3 - 2. \qquad (3.2)$$

But in many substances, for which $\alpha_0$ can be neglected, polarizability is inversely proportional to the specific spherical volume, in which dipoles are cooperating: $\alpha \sim (^4/_3\pi R^3)^{-1}$. Thus, it leads to such form of latent heat per particle:

$$\Delta U_b \approx (4\pi - 2) \kappa_B T_b, \qquad (3.3)$$

or, accordingly, for the molar energy of condensation,

$$\Lambda_b/T_b \sim 21 \text{ cal/mol K}, \qquad (3.4)$$

that just corresponds to (1.1), however now such expression can be already cited as the theorem. But for such generalization the necessary and sufficient conditions must be properly defined.

One more checking of usefulness of (3.1) as the general expression can be carried out as follows. The latent heat of transition into liquid state must tends to zero at approach to the critical temperature and pressure, therefore the general expression (3.1) would be rewritten for critical parameters as

$$\mathbf{d}^2/R_c^3 = 2\kappa_B T_c. \qquad (3.5)$$

Let's check up this relation, as an example, for water. As the critical temperature $T_c = 647$ K and if the dipole moment keeps its value $d = 1.855$ D, then $R_c = 2.68 \cdot 10^{-8}$ cm or the critical density must be of the order of $\rho_c = 0.372$ g/cm$^3$ at comparison with the experimental value $\rho_c^{(exp)} = 0.322$ g/cm$^3$.

But in general it is possible to assert that in spite of all approximations, including constancy of dipole moments, this result seems not unreasonable. It means that the offered theory allows the establishing of correspondence of critical parameters, at least, for certain substances. Notice that on the other hand, such relation can give possibility for investigation of possible changing of the dipole and other moments of molecules with changing of extensive parameters and for a certain tentative estimation of such changes.

Possible deviations from (3.4) in the scope of conducted discussion are evident. They can occur at approaching of temperature to the critical one, for particles of big sizes, with change of the dipole moment of particles at transition from one phase into another (usually $|\Delta\mathbf{d}| \ll |\mathbf{d}|$), by virtue of difference of a mirror dipole in dielectrics with the inducement dipole (difference is definable by values of dielectric susceptibility of both substances, e.g. [22]), in dependence on the form of condensate surface. All these changes, in principle, need to be taken into account directly in (3.1). More generally we can not assert that the interaction of complex molecules or clusters with condensable surface can be restrictively considered via influence of only one single electromagnetic moment.

It must be noted that the similar rules for processes of solidification of liquids are considerably less definite and practicable. The most simple of them assert that $\Lambda_m/T_m \approx 2.5 \pm 0.5$ cal/mole K for simple substances, $\Lambda_m/T_m \approx 6 \pm 1$ cal/mole K for inorganic molecules and $\Lambda_m/T_m \approx 13 \pm 3$ cal/mole K for organic ones. The consideration of these processes through proposed transitive radiation of own and induced momenta requires a revealing of "mirror" momenta inside the particles. Therefore their consideration should be much more complicated or would have less definite and more restricted heuristic significance. On the other hand their consideration is possible on the base of relation deducted in [7], the Appendix, via assumptions of frequencies and width of emitting levels responsible for phase transitions.

## 4. FREQUENCIES OF PHASE EMISSION

The general expressions, such as (2.6), demonstrate that the generated latent heat can be removal by radiation, but they do not determine its spectra. Therefore the peculiarities of possible energy emitters must be discussed. It is evident that the removal of energy by photons emission requires the existence of suitable levels in both phases, but we do not discuss here this quantum requirement.

In the general sense, without consideration of details of interactions and in view of much distinction of levels and conditions of constituents, there can be such opportunities:

1. If radiation has one-photon character,

$$\hbar\omega_1 = \Lambda/N_A \quad \text{or} \quad \lambda_1 = 120/\Lambda; \qquad (4.1)$$

where $\Lambda$ is expressed in kJ/mole, and lengths of waves in mcm, $N_A$ is the Avogadro number.

2. The n-photon transition with equal frequencies:

$$\hbar\omega_n = \Lambda/nN_A \quad \text{or} \quad \lambda_n = 120n/\Lambda. \qquad (4.2)$$

Here $n$ can be connected with the number of formed bonds in condensate, i.e. it can be of order of coordination number: n = 1, 2, 3, … . The singularities of spectra of condensate media can describe the number of bonds, i.e. their structure for simple substances at least. For monatomic substances the two-photon emission in the $^1S_0$ state with unchanging symmetry of systems are most probable.

3. In the case of more complex molecules transitions with quanta of various frequencies is possible, e.g. for two-photon case

$$\hbar\omega_1 = \hbar\omega' + \hbar\omega'' \quad \text{or} \quad \frac{\lambda' \cdot \lambda''}{\lambda' + \lambda''} = \lambda_1 \equiv \frac{120}{\Lambda}. \qquad (4.3)$$

More complicated combinations of frequencies can be similarly considered.

4. Dimers and more complicate formations, clusters, can be examined as single particles and if bound energies of atoms/molecules in them are small enough, wavelengths of radiation for a cluster of M particles will be of the type:

$$\lambda_1^{(M)} \approx 120/M\Lambda, \quad \lambda_n^{(M)} \approx 120n/M\Lambda \qquad (4.4)$$

(Here uncertainties can be connected with differentiation of bonds at formation of little clusters).

5. The more complicated situation with aggregation of M particles and additionally q = 1, 2, 3, …, m (m is the maximal coordination number) bonds can not be a priori excluded. It will mean the emission of (M + q/m) times latent energy:

$$\lambda_{n,m}^{(M)} \approx \frac{120n}{\Lambda}/(M + q/m) \qquad (4.5)$$

Such expression can be connected with appearing of certain bonds even at a stage of cluster formation.

The removal of all generated bond energy by one photon (the dipole radiation) seems the most simple and just it was assumed in the initial papers [10]. But the consideration in preceding section demonstrates that such process can not be the general feature. So, the entering of particles with dipole moments may be considered as formation and subsequent neutralization of the system of anti-parallel dipoles (the anti-parallelism is marked by arrows):

$$\vec{d}(x, y, \Delta z) + \overleftarrow{d}(x, y, -\Delta z) \to A(x, y, 0) + \gamma_1 + \gamma_2 + \ldots, \qquad (4.6)$$

where through $A(x, y, 0)$ a neutral "formation", which already had emitted the energy of transition and had been included in substance, is designated. Notice that the interaction between these dipoles takes place in the near field [23], i.e. can be instantaneous [24], the transferring of excitations between anti-parallel dipoles in near field was supervised experimentally [25].

This dipoles system has the positive charge parity and therefore can be neutralized, without change of symmetry, by emission of two photons, probably of equal frequencies, in the $^1S_0$ state with $\Delta J = 0$, i.e. by (4.2) with n = 2. The decay rate of dipoles system with emission of two photons, if they are entangled, coincides with the decay rate of single dipole and can be very roughly estimated via the usual expression for such radiation:

$$1/\tau_E \sim \alpha(ka)^2 \omega = (2\pi)^3 d^2/\hbar\lambda^3. \qquad (4.7)$$

With the characteristic values d ~ 1 D; $\lambda$ ~ 6 mcm for water it leads to $\tau_E \sim 10^{-3}$ sec that leads, for example, to drops growth in the vapor phase with the maximal velocity of the order from mcm till tens mcm per second. Such estimation seems non-inconsistent.

This process is equivalent to the emission of one scalar particle and in an accord to the Goldstone theorem can lead to changing of the symmetry of system, e.g. at crystallization in the vapor phase. As one example here can be added that the spectrum of water contains the lines of order of 6 mcm, two photons of which correspond to the generated latent heat at vapor condensation. (Corresponding lines can be determined and shown for many other substances [11].) The calculations of their probability, i.e. the determination of processes duration, can be carried out by a close analogy with the positronium decay.

The two-photon, the very often, removal of energy generated at the first-order phase transitions may be explicable by such reasons also. The van der Waals interaction is describable in QED by the two-photon exchange [26] and the imaginary part of its matrix element is proportional, in the spirit of optical theorem, to probability of emission of two photons. Therefore such removal of latent energy corresponds to the van der Waals intermolecular bonds.

# 5. EXPERIMENTAL RESULTS FOR ALKALI HALIDES AND SAPPHIRE (IR-TRANSPARENT)

Initially, the investigation of IR emission spectra during crystallization in [27, 28] was performed on an apparatus consisting of a resistance furnace, in which a platinum or alundum crucible without a lid was placed, and a one-beam IR spectrophotometer with remote globar. The standard spectrometer was modified for the recording of the rapid processes. The radiation from the melt-crystallizing substance system was directed to the spectrophotometer by a metal mirror.

Crystallization led to formation of a polycrystalline bar within a period ranging from a few tens of seconds to an hour in dependence on the type of cooling. During crystallization of some substances (LiF, NaCl, NaBr, NaI, KCl, KBr, KI, $PbCl_2$) the recorded curves exhibited not only the thermal radiation, but also additional peaks corresponding the wave length $\lambda \approx 2 \div 4$ mcm. The Fig. 1a from [28] presents a set of five consequent spectra in relative units recorded during LiF cooling. The curve of the melt temperature change during the experiment is presented on the same figure. The less inclined part of the curve corresponds to melt crystallization. We can conclude that the appearance of peak precedes the start of crystallization that can be explained by cluster formation in the melt. The peak does not disappear immediately after visual completion of crystallization. The time taken for its disappearing depends on the rate of crystallization and ranges from a few seconds to a period of from two to five minutes.

The value of the peak 3 is higher than that of peak 2 in spite of the decreasing system temperature and the resultant decrease in the background radiation. The decreasing temperature is accompanied by an increase of the crystallization rate, so the intensity of characteristic radiation increases with the crystallization rate.

In [27] the energy corresponding to the spectrum was found to be higher than latent energy of crystallization and this fact was explained by the hypotheses of Jackson and Chalmers [29] concerning crystallization. Their model postulates that crystallization is an activated process with energy of activation equal to that of self-diffusion or viscous flow.

In [28] the experiments were mainly carried out with LiF because of its low hygroscopicity (later LiF and KBr were found to be the most acceptable for the experiments because of their low melt viscosity). The spectra structures and peak positions from [28] shown in Fig.1a are similar to those from [27]. They are rather complicated because of lines of adsorption of atmospheric water vapor (3.4 mcm), atmospheric $CO_2$ (3.95 mcm and 4.35 mcm) and bonded with crystallographic LiF structure $OH^-$ groups (2.7 mcm). Later, in [28] the parasitic bands were eliminated to allow more precise determination of the peak wavelengths. For this, first of all, the experiments were carried out in the atmospheres of Ar or $N_2$. The results are presented in the Fig. 1b.

The second step in improving detection of the pure characteristic spectrum consisted of using a two-beam spectrometer with compensation of the heat radiation of the crucible and salt melt. This allowed to more accurate determination of the peak positions and sometime to resolve its fine structures. The resulting LiF peak presented in Fig.2 can be interpreted as a superposition of four peaks with $\lambda_1 = 2.80$ mcm; $\lambda_2 = 3.45$ mcm; $\lambda_3 = 4.05$ mcm; $\lambda_4 = 4.35$ mcm. The position of the last peak exactly corresponds to the latent energy of crystallization.

In spite of our efforts (including a fast crystallization on a cooled copper substrate), we could not resolve the fine structure of crystallization radiation for other investigated alkali halides. Only the main peaks corresponding to the latent heats of phase transitions were detected. So, in this manner we could relatively easily observe the crystallization radiation of low-viscosity melts which are transparent in the near IR region. For more viscous melts, such as KCl, it was difficult to observe and even more to separate pure crystallization radiations. Probably, a reason of this has been a glass formation during fast solidification.

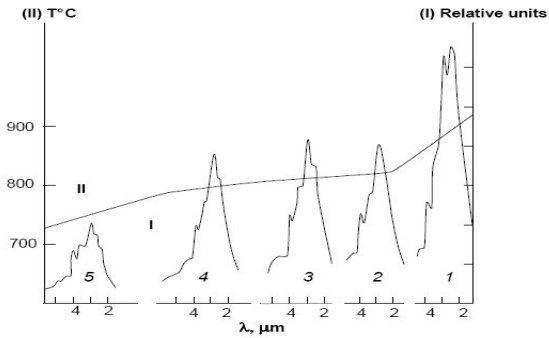

*Fig.1a*

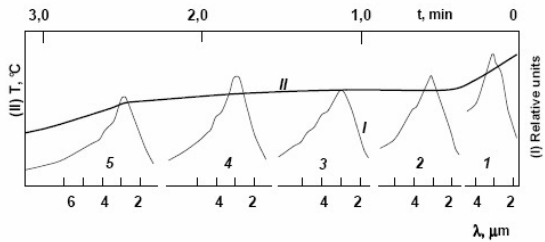

*Fig. 1b.*

***Fig. 1*** *(from [28]): a) Open air environment: b) Argon environment: Thermogram (II) and radiation spectra (I) of crystallizing LiF: 1 - Spectrum of superheating melt; 2, 3, 4 - Spectra of crystallizing melt; 5 – Spectrum of crystallized substance.*

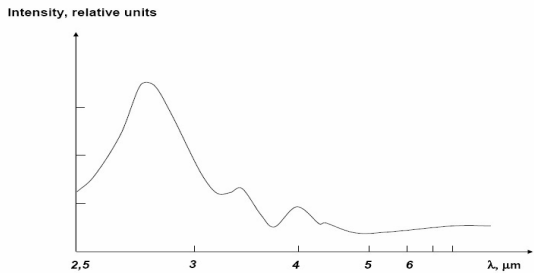

***Fig.2*** *(from [28]). Characteristic radiation spectrum of crystallizing LiF melt.*

In [30], we described our experiments concerning recording of the characteristic spectra during sapphire ($Al_2O_3$) crystallization from the melt. Sapphire is transparent over a wide range of the spectrum from infra-red to ultra-violet [31]. The substance is interesting on account of its low entropy of fusion, corresponding to slight structure differences between the crystal and the melt, in contrast with semiconductors and alkali halide crystals. We measured the emission spectra of solid, liquid, and crystallizing sapphire in the spectral region from 0.7 to 2.5 mcm at the temperatures between 2100°C and 1600°C. Heating and melting were affected in a vacuum with the aid of a high-frequency apparatus fitted with optical windows of fluorite or fused quartz. Other details of apparatus are given in [30].

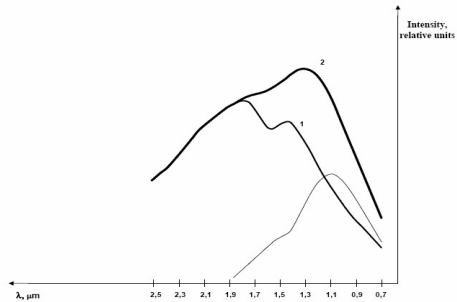

*Fig.3* (from [30]). *Radiation spectra of sapphire: Differential spectrum of radiation flare (dashed curve). 1. Emission spectrum of cooled sapphire spectrum first heated to the temperature less than 2050°C. 2. Emission spectrum of crystallizing melt.*

The main result is presented in Fig. 3. From the spectrum of crystallizing melt 2, we subtracted the emission spectrum of crystalline sapphire first heated to the temperature less than the melting temperature 2050°C. Thus the crystallization of sapphire from the melt is accompanied by the appearance of a band of IR radiation (additional to the thermal radiation) at the wavelength of 1.1 mcm, corresponding to the heat of fusion of sapphire 26 kcal/mol.

So it must be concluded that the energetic spectrum of LiF contain four bands: $\lambda_1 = 2.80$ mcm; $\lambda_2 = 3.45$ mcm; $\lambda_3 = 4.05$ mcm; $\lambda_4 = 4.35$ mcm. With the accuracy 7% the energies of the set of peaks correspond to the formula (4.5): $\Lambda$ - a latent heat of phase transition for one particle (n = 1) for LiF is equal 26.4 kJ/mol; m, a coordination number for LiF, is equal 6; (M, q) = 0, 1, 2,..m. The $\lambda_1$ corresponds (M, q) = (1,3) ; $\lambda_2$ : (M, q) = (1, 2); $\lambda_3$ : (M, q) = (1, 1); $\lambda_4$ : (M, q) = (1, 0).

Furthermore as in [28] we used more accurate technique with respect to [27] and determined more precise band positions for all alkali halides investigated. We now believe that our interpretation of the process as described in [27] probably has to be corrected. We suggested that the energy of radiation is the latent energy of phase transition plus the activation energy. But the majority of particles lose the energy of activation for transition of the potential barrier. So, the liberated energy mainly has to be equal to the latent energy of phase transition. Nevertheless, we don't exclude that some particles can liberate the energy that is equal to the sum of the latent energy and the additional kinetic energy that is the most probable for the excited particles in the gas phase.

## 6. WATER PHASE TRANSITION

Here we present some experiments concerned with the infrared radiation appearance during water phase transitions.

In [32] an infrared line scan photo camera has been developed that scans an object simultaneously with three separate spectral regions and produces an image of the object as a color photograph. The three spectral regions are 0.5 ÷1.0 mcm, 3.0 ÷5.5 mcm and 8 ÷ 14 mcm. Each of the infrared spectral regions is rendered in one of the primary colors: blue, green and red respectively. As a result, the color of objects in a picture indicates their temperature and also their reflective and emissive properties. The pictures presented in the paper allow to conclude that the authors found the infrared sources of the range 8 ÷14 mcm radiations in the atmosphere that can be concerned neither with temperature radiation nor reflective one. These sources were the bottom sides of cumulus clouds with the temperature –5ºC and the rising warm air saturated with water vapor.

The experiments [33] were carried out in the installation including a vessel with boiling water, a cooled glass surface for the vapor condensation, and a sensitive system of the infrared radiation recording. There was observed an anomalous increase in the infrared radiation intensity from the boundary of glass surface, i.e. from the condensed vapor. The intensity was increased with the condensation rate increasing. At the range of 1÷ 4 mcm the integral intensity was four times bigger the Plank's radiation. Two main emission bands were seen at 2.10 mcm and 1.54 mcm wavelengths (Fig. 4). The intensity of both bands exceeded the background radiation by a factor of ten. Probably, third and forth peaks with $\lambda$ = 2.5 mcm and $\lambda$ = 3.2 mcm which were not mentioned by the authors can be recorded on the curve Fig.4.

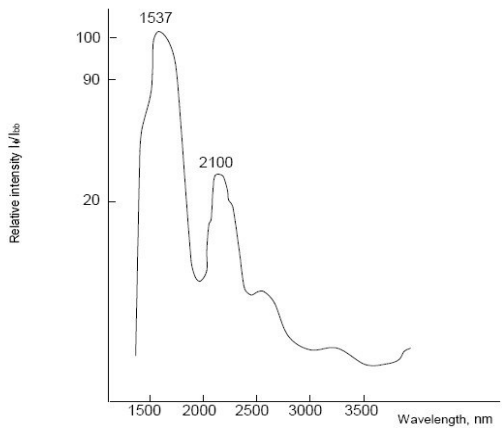

***Fig.4*** *(from [31]). Relative intensity of the phase transition luminescence from water; $I_w$, intensity from the boiling water; $I_{bb}$, intensity from the black body.*

In [34], at observation of sonaro-luminescence, the author used a spectrometer with a thermocouple detector and a Lead Sulphide photometer. The peak of emission with wavelength 1.05 mcm was detected. Probably, second peak with $\lambda$ = 0.9 mcm, which was not mentioned by the author, can be recorded on this curve.

In our experiments [35] characteristic radiation was recorded during water vapor condensation and water crystallization in the closed chamber. The temperature was decreased by adiabatic expansion. The radiation was led out from the chamber through the Ge window. A system of filters and mirrors cut specially selected infrared ranges. Radiation was detected by a bolometer. This technique allowed to confirm that the characteristic radiation with a maximum in the range 4 ÷ 8 mcm for the first process and in the range of 28 ÷40 mcm for the second process is recorded. The radiation intensity many times exceeded the background Plank's intensity for the same temperature.

Now one can refer to infrared Earth's images from the Space by NASA (March 23, 2002 [36]): the orbiting GOES 8 satellite's multi-channel imager produces images of upper troposphere at the infrared wavelength of 6.7 mcm. Bright regions on it correspond to high concentrations of water vapor, while dark spots are relatively dry areas.

Let's discuss these results. If each molecular transition from one phase to another produces two photons, the usual magnitudes of latent heat must lead to such values: 1). For water condensation (for instance, drop formation from the vapor cloud formation in the atmosphere), the main peak must correspond to $\lambda_{1w}= 5.80$mcm; 2). For water crystallization (water drops freezing at hail formation in the atmosphere) $\lambda_{2w}= 39.24$mcm; 3). For crystallization from vapor phase (snow particle formation) $\lambda_{3w} = 5.06$mcm. With taking into account the Doppler's broadening of characteristic radiation, the following ranges for $\lambda_{1w}$ and $\lambda_{2w}$ can be estimated: $\lambda_{1w}= (5.2 \div 6.4)$ mcm and $\lambda_{2w}= (34 \div 42)$ mcm.

Thus, in [35], the increased intensity of radiation was found in the ranges $\lambda_{1w}$ and $\lambda_{2w}$. The experimental results from [32] correspond to $\lambda_{1w}$ with taking in account of low pressure of vapor condensation. In [36] near $\lambda_{1w}$ range radiation was recorded.

In [33] the positions of all from four peaks correspond to emitted energy that exceeds the phase transition one. The peaks of radiation on 1.5 mcm, 2.1 mcm and, apparently, on 2.5 mcm and 3.2 mcm, with taking into account the relation (4.4), can be attributed to condensation of water's dimers and higher molecular complexes with n = 2 and m = 4, 3, 2.

Some features of the sonaro-luminescence phenomena [34] can be described via emission of latent energy with further multiphoton composition of their frequencies or the peak of emission with $\lambda \sim 0.9$ mcm and 1.05 mcm correspond (4.4) with m = 6 and 5 consequently (water is opaque for lower frequencies).

## 7. EXPERIMENTAL RESULTS FOR Te, Ge AND METALS (IR NONTRANSPARENT)

A differential technique of radiation recording has been used to study the crystallization of Te [37]. Ten grams of tellurium were sealed under a vacuum into a quartz ampoule. A similar quartz ampoule was filled with a powder of graphite and metal shaving. Both ampoules were inserted into a furnace. The temperature in the ampoules was measured by thermocouples. Radiation beams from the ampoule bottoms in the optical range $2 \div 4$ mcm (the transparency window of the system) were collected on two PMT-detectors, the difference of electric signals was detected by the oscilloscope. Thus, the detecting system allowed recording very fast processes. The crystallization began with an undercooling of ~100 ºC. A very intensive radiation impulse of 1.5 ms duration was recorded at the very beginning of every from 20 crystallizations.

The latent energy of Te melt crystallization is 17.5 kJ/mol. For one photon characteristic emission, this energy corresponds to $\lambda = 6.8$mcm. The radiation recorded in [37] corresponds to the second harmonic. The transparency window of the system (2 – 4)mcm doesn't allow observation of other lower or higher harmonics, if even they exist.

In [38] the registration of emission spectra of cooled germanium, previously melted, was carried out with a high-speed spectroscopy technique. The crystallization of germanium was accompanied by sharp increasing of its radiation in near IR region of a spectrum. The most effect was recorded for a high optical quality germanium. There is only indication that a radiation of increased intensity corresponds to a near IR range.

Luminescence has been recorded while copper and aluminum wires have been exploded by a high electric current [39]. A luminescence bands with peak intensity near ~ 1.5 mcm, 0.85 mcm, 0.80 mcm and 0.70 mcm were recorded for copper wires and near ~ 0.43 mcm and 0.36 mcm for aluminum wires.

The latent energy of Cu melt crystallization is 13 kJ/mol. For one photon characteristic emission, the wavelength would be $\lambda_1 = 9.2$ mcm. The 1.5 mcm radiation detected in [36] corresponds to sixth harmonic. The system of recording used there did not allow detection of lower harmonics. At the same time, some non-identified bands 0.70 mcm, 0.80 mcm, 0.85 mcm could correspond to higher harmonics. We also don't exclude that condensation of the metal vapor after the wire explosion could produce the characteristic radiation. The latent energy of Cu vapor condensation is 304 kJ/mol. One of the unidentified bands may correspond to the two photon characteristic emission at $\lambda_3 = 0.80$ mcm.

For aluminum wire explosion the situation is more definite: in accordance with (4.2) $\lambda \sim 0.43$ mcm corresponds to the latent heat of condensation $L_b = 68$ kcal/mole and $\lambda \sim 0.36$ mcm – for sublimation $L_{sub} = 80$ kcal/mole

**CONCLUSIONS**

Let us enumerate the results and some further perspectives.

1. Characteristic radiation corresponding to the first order phase transitions in more ordered states exists and can remove out the latent energy of transition (its parts at least).

2. This phenomenon, for the first time, ascertains the internal conformity between the thermodynamic and spectroscopic magnitudes. It is describable by electromagnetic moments of particles and forming bonds.

3. The investigated phenomena demonstrate possibilities and perspectives of "the spectroscopy of phase transitive radiation". The significance of such spectroscopic researches consists in particular in possibility of spectroscopic analyses of clusters and their structure. It can be added that a peculiar contribution into the spectra would be introduced by clusters: so the water clusters contain bigger dipole moments [40] and therefore their condensation would lead to higher frequencies.

4. Such investigations can have especial sense, in particular, in astrophysics, where are known many objects with big excesses of near and mid-IR radiation (e.g. [11], more recent supervision data e.g. in [41]).

5. The existence of phase radiation allows control of transition processes. New systems of crystallization process regulation on the basis of radiation recording can be developed. As further perspectives the management of processes of solids formation can be, probably, considered.

6. According to the general principles of quantum theory the presence of described spontaneous transitions should lead to an opportunity of stimulation of such transitions, for example crystallization, by irradiation of the substance, close to the temperature of crystallization, at a resonant frequency corresponding to the particular path of reaction. Such opportunities, which until now have not been investigated experimentally, should stimulate new effects. For instance, fog formation may be observed as a result of atmosphere irradiation by the characteristic radiation. Sometimes, the crystallizing system would have to be protected against the characteristic radiation to decrease the probability of parasitic nucleation.

7. A program could also be suggested to use this effect in the study of climate problems. Atmospheric radiation is one of the key subjects of the atmospheric sciences, linking the fields of chemistry, aerosol and cloud physics, and thermodynamics to global climate and climate change. The Atmospheric Radiation Measurement (ARM) Program was created to help resolve scientific uncertainties related to global climate change, with a specific focus on the crucial role of clouds and their influence on radiative feedback processes in the atmosphere. But in spite of the huge scale of the program, including 50 countries, attentive analysis of the program shows that it could be improved. A measurement of intensity of characteristic radiation accompanying water vapor condensation with clouds formation, as well as water drops crystallization with hail and snow clouds formation, could provide significant information concerning cloud formation, as well as the energy balance in atmosphere. We could also use these radiation measurements to know if there is water in the atmosphere of other planets, for instance in the Mars' atmosphere.

8. In [33] the recorded energy could be estimated as 3 – 5% of full phase transition energy. The problem of characteristic radiation yield has not been solved yet in the frame of offered theory. It is evident that the yield, first of all, depends on the transparency of both phases. But here the transparency problem is a very specific one. Indeed, our system contains the main level and the exited one. So, it can work as an amplifier for the characteristic radiation and, as a result, to be transparent for it. This is the only explanation for the detection of the characteristic radiation of phase transitions for water, ice, Te, Al and Cu. All these substances are non-transparent for infrared radiation. Practically, homogeneity and geometry of the system under consideration have to be very important.

9. As it was mentioned in [42] the IR characteristic radiation corresponding to the crystallization heat can ensure crystallization heat removal from the phase-transition boundary. It has to be taken into account for Stephan's problem solutions.

10. The considered processes of the removal of main part of latent heat do not exclude possible partial processes connected with features of different substances and their constituents. So at condensation of water molecules there can exist low frequencies (radio) emission, energy of which is very small [43].

Let's briefly notice certain other perspectives

The used methods can be applied to investigation of electromagnetic interactions between neutral formations and particles that are now intensively researched, e.g. [44]. It allows the specification of these relations for different types of substances and different external influences (we pay attention, in particular, on specifications of the method of mirror charge [45]). Besides, the deduced relations ascertain correlations of certain, at least, critical parameters of substances and/or changes of their internal characteristics.

The considered picture of phase transitions with the specific non-Planck radiation shows the essential distinctions between quantum processes of condensation and evaporation: the energy necessary for evaporation can be, basically, collected by thermal fluctuations, etc. Hence though these processes thermodynamically are reciprocally related, their comparison as reversible ones becomes impossible, and it can explain difficulties of many preceding kinetic approaches [4, 5].

It must be noted that the similar radiation (mainly in the infrared region) at processes with neutral particles must accompany, apart from known photochemical reactions, some other phenomena of physical and chemical adsorption. So, it can be interesting to consider on this base, among others, processes of dissolution, since at dissolution of many substances the dipole moments of molecules can be distinctly changed: e.g. $CH_3Cl$ has $d_{gas}$ = 1.86 D and at dissolution in benzyl this moment varies on $\Delta d \sim -0.24$ D; molecules of HCN have, accordingly, $d_{gas}$ = 2.93 D and $\Delta d \sim -0.34$ D. Therefore processes of their dissolution also should be accompanied by generation of the latent heat of dissolution, which, under condition of medium transparency, can be converting into a characteristic radiation.

Here must be noticed that some close effects are described as crystalloluminescence (CL). The effects of CL investigated for a long time, are including, apparently, displays of several very different phenomena: crystallization including nucleation processes, accumulation of transitions energy on the introduced radiative centers, polymorphic transformations, chemiluminescence, luminescence with fractures and cracks formation, triboluminescence, etc. (e.g. [46]). But they do not include vapor condensation into liquid. Such mixture complicates their examination and we try to separate from possible sources of radiation such ones that are caused exclusively by phase transitions of the first kind.

Let's note, that it is possible to attribute to an investigated class of the phenomena also a radio emission, accompanying phase transitions in hydrocarbons [47]. Also such processes as formation of the double electric layers, the original "phases" (they can be considered as occurrence of system of dipoles) at which such radio emission has been fixed [48] can be attributed to analogical processes.

As was indicated above, in our experiments with alkali halides and sapphire, the time of appearance of the peak does not always coincide with the time at which the crystallization process is recorded. If bonds of various strengths in the crystal are broken or formed during the phase transition, the phase transition occurs smoothly (sometimes with peculiarities) in some temperature interval. This temperature interval can be determined by infrared spectroscopy. For instance, this technique was used in [49] for determination of the temperature interval of cyclo-hexane phase transition. It was not a crystallization process, but it is interesting as another example of radiation accompanying phase transition. A cyclo-hexane crystal transformed, at T = 186 K, from a plastic (high-temperature) to an anisotropic (low-temperature) modification. The phase transition at this temperature is accompanied by radiation of the frequency 55 cm$^{-1}$. The optical energy emitted by a crystal is a part of the electromagnetic component of polariton energy (The polariton energy is equal to the energy of a transverse electromagnetic wave with a small admixture of the mechanical vibration energy of a dipole). The emission appears at the temperature about 190 K, grows to its maximum value at the temperature 186 K and smoothly decreases down to the temperature 183 K.

The received results cannot be considered, certainly, as the absolutely exact. It is enough to note that the Casimir forces are not taken into account. Notice also possible complications of the proposed model, connected with consideration of higher moments, omitted for simplicity in this paper. However apart from considered processes can be such, in which only part of latent heat is immediately converted into radiation emission by one act, the process can go in the form of multi-step adsorption (physical and chemical) with partial removal of latent heats, i.e. via sequential steps of phase transition. The possibility of dimers and more complicated formations of gas particles before condensation would be also noticeable in the spectra of phase radiation. On the other hand molecules of a tail of the Maxwell distribution can peculate their latent energy via collisions close to condensate surface and therefore can be accumulated with emission of lower frequencies for which proper levels can exist.

## ACKNOWLEDGEMENTS

The first author is greatly indebted to critical remarks of M. Ya. Amusia, Y. B. Levinson, M. I. Molotsky, I. I. Royzen, G. M. Rubinstein and V. R. Shaginian. Second author would like to thank Mr. H. Bates and Dr. V. Ouspenski for fruitful discussions.

**REFERENCES**
[1]. M. E. Perel'man and V. A. Tatartchenko. Phys. Lett. A, doi:10.1016/j.physleta.2007.11.056
[2]. J. Scwinger, L. L. De Raad and K. A. Milton. Ann. Phys. (NY), **115**, 1, 388 (1978).
[3]. P. W. Milonni and P. B. Lerner. Phys.Rev. A, **46**, 1185 (1992); P. W. Milonni. *The Quantum Vacuum*. L.: Academic, 1994. .
[4]. M. Volmer. *Kinetik der Phasenbildung*. Dresden: 1939; R. F. Stricland-Constable. *Kinetics and Mechanism of Crystallization*. L.: Academic Press, 1968..

[5]. E. C. Baly. Phil. Mag., **40**, 15 (1920); Ch. Schaefer und F. Matossi. *Das Ultrarote Spektrum*, B.: Springer, 1930, p. 284.
[6]. C. Kittel. *Quantum Theory of Solids*. NY, Wiley, 1963, p. 320.
[7]. M. E. Perel'man. Phil. Mag., **87**, 3129 (2007).
[8]. J. Goldstone. N.Cim., **19**, 154 (1961).
[9]. K. J. Gaffney, et al. Phys. Rev. Lett., **95**, 127501 (2005).
[10]. M. E. Perel'man. Phys.Lett. A, **32**, 411 (1971); Sov.Phys.-Doklady, **203**, 1030 (1972); **214**, 539 (1974).
[11]. M. E. Perel'man. Astrophysics, **17**, 213 (1981).
[12]. F. T. Trouton. Phil. Mag., **18**, 54 (1884).
[13]. L. K. Nash. J. Chem. Educ., **61**, 981 (1984); J. Wisniak. Chem. Educator, **6**, 55 (2001).
[14]. V. L. Ginzburg and I. M. Frank. Zh. Eksp.Teor.Fiz., **16**, 15 (1946).
[15]. V. L. Ginzburg and V. N. Tsytovich. Phys. Rep., **49**, 1, 1979; *Transient Radiation and Transient Scattering*. Taylor & Francis, 1990.
[16]. V.N. Bogomolov, E.K. Kudinov, Yu.A. Firsov, Soviet Solid State Physics 14 (1972) 2075.
[17]. A. Zangwill. *Physics at Surfaces*. Cambridge, 1988.
[18]. M. E. Perel'man and G. M. Rubinstein. Sov. Phys. Doklady **17**, 352 (1972); In: *New Research on Lasers and Electro Optics* (W.T. Arkin, Ed.). NY: Nova Sc. Publ., 2006, pp. 229-267.
[19]. S.A. Sall', A.P. Smirnov,: Technical Physics, **45** (2000) 84958.
[20]. J. O. Hirschfelder, Ch.F.Curtiss and R.B.Bird. *Molecular Theory of Gases and Liquids*. J.Wiley, 1954.
[21]. W. F. Braun. *Dielectrics*. In: *Handbuch der Physik* (Heraus. S.Flügge). **XVII**, 1. Springer, 1956.
[22]. J. D. Jackson. *Classical Electrodynamics*. J.Wiley, 1962.
[23]. M. E. Perel'man. Ann. Phys. (Leipzig), (Leipzig), **16**, 305 (2007).
[24]. M. E. Perel'man. Ann. Phys. (Leipzig), **14**, 733 (2005).
[25]. T. Kawazoe et al. e-J. Surf. Sci. Nanotech., **3**, 74 (2005) and references therein.
[26]. V. Berestetski, E. Lifshitz and L. Pitaevski. *Relativistic Quantum Theory*. Oxford: Pergamon, 1971.
[27]. V.A.Tatarchenko, Soviet Physics - Crystallography **24,** 238 (1979).
[28]. L.M. Umarov, V.A. Tatarchenko, Soviet Physics - Crystallography **29**, 670 (1984).
[29]. B. Chalmers, *Principles of Solidification.* NY, Wiley & Sons, 1966, p. 42.
[30]. V.A. Tatarchenko, L.M. Umarov, Soviet Physics - Crystallography **25**, 748 (1980).
[31]. V.A.Tatartchenko, *Sapphire crystal growth and applications*. In: P. Capper (ed.), *Bulk crystal growth of electronic, optical and optoelectronic materials.* London, Wiley & Sons, 2005, p. 337
[32]. L.W. Nichols, J. Lamar, Applied Optics **7**, 1757 (1968).
[33]. W.R. Potter and J.G. Hoffman, Infrared Physics **8**, 265 (1968).
[34]. M. Ayad, Infrared Physics **11**, 249 (1971).
[35]. A.N. Mestvirishvili, J.G. Directovich, S.I. Grigoriev, M.E. Perel'man, Phys. Lett. **60**A (1977) 143.
[36]. F. Hasler et al, http://antwrp.gsfc.nasa.gov/apod/ap020323.html.
[37]. I.V. Nikolaev, B.I. Kidyarov, A.P. Kozharo, Abstracts of III International conference "Kinetics and mechanisms of crystallization", Ivanovo, Russia, 2004, p. 34.
[38]. O.D.Dmitrievskiy, S.A.Sall', www.physical-congress.spb.ru/download/cong04(2003).
[39]. M.I.Molotsky, B.P. Peregud, Sov.Tech. Phys. **26** (1981) 369. K.B. Abramova, B.P. Peregud, Yu.N. Perunov, V.A. Reingold, I.P. Stcherbakov, Optics and Spectroscopy, **58** (1985) 809.
[40]. R. Moro et al., Phys. Rev. Lett., **97**, 123401 (2006); R. Bukowski, et al. Science, **315**, 1249 (2007).
[41]. Ch. J. Lada and E. A. Lada. Ann.Rev.Astron.Astrophys. **41**, 57 (2003); C Dijkstra et al. Bull. Am. Astronom. Soc., **37** #2 [8.02] 2005.
[42]. V.A.Tatartchenko, *Shaped crystal growth* (Kluwer Academic Publishers, London, 1993) p. 3
[43]. M. E. Perel'man, G. M. Rubinstein and V. A. Tatartchenko. In: arXiv:0712.2564v1
[44]. M. Ya. Amusia. Rad. Phys. Chem., **75**, 1232 (2006) and references therein.
[45]. P. Bushev et al. Phys. Rev. Lett., **92**, 223602 (2004).
[46]. M. Barsanti and F. Maccarrone *Riv. Nuovo Cimento* (1991) # 14; B P Chandra, P K Verma and M H Ansari. J. Phys. C. **9**, 7675 (1997).
[47]. I.Sugawara, Y.Tabata. Chem.Phys.Lett., **41**, 357 (1976).
[48]. N.G.Khatiashvili, M.E.Perel'man. Phys. Earth & Plan. Int., **57**, 177 (1989) and references therein.
[49]. E.A. Vinogradov, G.N. Zhizhin, Solid State Physics **18**, 2026 (1976).